\newcount\Comments
\documentclass[preprint,twocolumn]{aastex701}

\usepackage[T1]{fontenc}
\usepackage{lmodern}
\usepackage{microtype}
\usepackage{amsmath}
\usepackage{graphicx}
\usepackage{latexsym}
\usepackage{bm}
\usepackage[english]{babel}
\usepackage{color}
\usepackage{natbib}
\usepackage{hyperref}
\usepackage{soul}
\usepackage{xcolor}
\usepackage{graphicx}
\usepackage{amsmath}
\usepackage{booktabs}
\usepackage{bm}
\usepackage{makecell}
\usepackage{placeins}

\definecolor{purple}{rgb}{0.6,0,1}
\definecolor{pink}{rgb}{1.0,0.41,0.71}
\definecolor{orange}{rgb}{1,0.6,0}

\newcommand{\kibitz}[2]{\ifnum\Comments=1\textcolor{#1}{#2}\fi}

\begin{document}

    \title{SKYSURF-IR: The Effect of Source Crowding on Faint Galaxy Completeness in JADES}

%
\author[0009-0002-9671-7606,gname=Kendall,sname=Johnston]{Kendall Johnston}
\affiliation{School of Earth and Space Exploration, Arizona State University, Tempe, AZ 85287-6004, USA}
\email{replace-kendall-email@example.invalid}

\author{Logan R. Conrad}
\affiliation{School of Earth and Space Exploration, Arizona State University, Tempe, AZ 85287-6004, USA}
\email{replace-logan-email@example.invalid}

\author[0000-0003-0238-8806,gname=Darby,sname=Kramer]{Darby Kramer}
\affiliation{School of Earth and Space Exploration, Arizona State University, Tempe, AZ 85287-6004, USA}
\email{replace-darby-email@example.invalid}

\author[0000-0001-6650-2853]{Timothy Carleton}
\affiliation{School of Earth and Space Exploration, Arizona State University, Tempe, AZ 85287-6004, USA}
\email{tmcarlet@asu.edu}

\author[0009-0009-7696-9137]{Kamran Oskouie}
\affiliation{School of Earth and Space Exploration, Arizona State University, Tempe, AZ 85287-6004, USA}
\email{koskouie@asu.edu}

\author[0000-0002-9984-4937]{Rachel Honor}
\affiliation{School of Earth and Space Exploration, Arizona State University, Tempe, AZ 85287-6004, USA}
\email{rchonor@asu.edu}

\author[0000-0002-6150-833X]{Rafael Ortiz III}
\affiliation{School of Earth and Space Exploration, Arizona State University, Tempe, AZ 85287-6004, USA}
\email{rortizii@asu.edu}

\author[0000-0003-3329-1337]{Seth H. Cohen}
\affiliation{School of Earth and Space Exploration, Arizona State University, Tempe, AZ 85287-6004, USA}
\email{seth.cohen@asu.edu}

\author[0000-0001-8156-6281]{Rogier A. Windhorst}
\affiliation{School of Earth and Space Exploration, Arizona State University, Tempe, AZ 85287-6004, USA}
\email{windhors@asu.edu}

\begin{abstract}
SKYSURF-IR is an archival project that uses the James Webb Space Telescope (JWST) to measure diffuse light in the near-infrared, as a follow-up to the Hubble Space Telescope's project SKYSURF \citep{Windhorst2022,Carleton2022}. SKYSURF concluded an underestimation of sky-surface brightness values \citep{Brien2023}. One cause could be an underprediction of the Integrated Galaxy Light (IGL), resulting from the extrapolated faint galaxy counts. In this paper, we therefore assess the effect of source crowding in JWST/NIRCam imaging of the JADES GOODS-S deep field by adding successive rotated copies of detected sources to the masked square images. In the NIRCam F150W filter, the $2\times$, $3\times$, and $4\times$ image realizations resulted in completeness levels of 79.0\%, 71.3\%, and 66.2\% of the galaxy counts expected from scaling the original catalog, respectively. The recovered counts drop farthest below the expectations at the faintest magnitudes, providing insight into how image crowding limits the completeness of the galaxy counts in this field, which in turn affect the estimates of the integrated galaxy light.
\end{abstract}

\section{Introduction}
Extragalactic background light (EBL) is the cumulative radiation that is produced by galaxies and accreting black holes over cosmic time, making it a fundamental constraint on the history of star formation, structure formation, and dust processing in the universe \citep{Driver2016,Mattila_Vaisanen2019,Driver2021}. In principle, the optical and near-infrared EBL can be estimated either from direct measurements of the diffuse sky brightness or by integrating the light from individually detected galaxies in deep-survey images. In practice, however, these two approaches have not always yielded consistent results \citep[see, e.g.,][for a review]{Driver2016}.

It is estimated that the cosmic infrared background (CIB)  is made up solely of distant galaxy light, though CIB estimates from direct and integrated measurements do not yield consistent results \citep{Driver2016,Mattila_Vaisanen2019,Koushan2021,Driver2021}. At the same time, recent work has shown that uncertainty remains in both direct measurements and infrared contributions from faint galaxies below the telescope detection limits. The Hubble Space Telescope (HST) SKYSURF project found that constraints on the diffuse near-infrared background vary with how foreground light is modeled \citep{Windhorst2022,Carleton2022}. New Horizons measurements renewed the debate over whether the optical extragalactic background light includes a possibly significant diffuse component \citep{Lauer2021}. Using JWST/NIRCam observations, the PEARLS survey measured both galaxy counts and integrated galaxy light over a wavelength range of $0.9$--$4.5\,\mu\mathrm{m}$ \citep{Windhorst2023}. Together, these studies motivate a closer re-examination of the integrated galaxy-count approach, particularly whether faint-source incompleteness could bias the total integrated light derived from deep imaging surveys.

One key mechanism that can reduce catalog completeness is source confusion. This can arise from large instrumental Point Spread Functions (PSFs) when fewer than 25--50 PSF FWHM beams are available in the sky to detect and disentangle the faintest sources. But it can also arise when the average faint object sizes become considerably larger than the PSF FWHM, as is the case for the excellent small PSF size and substantial depth reached in HST and JWST images \citep[see, e.g.,][]{Kramer2022}. As the HST and JWST observations reach a greater depth, the surface density of faint galaxies increases, and sources are more likely to start overlapping in projection due to their own angular sizes, rather than due to the small PSF FWHM of HST and JWST. Under these conditions, compact faint galaxies may be affected by both source confusion and instrumental limitations, causing them to either blend with nearby objects, become buried by nearby sources, or be missed entirely by standard detection methods. This issue has previously been investigated in detail through the deepest HST imaging in the Hubble Ultra Deep Field (HUDF), where \cite{Kramer2022} found that crowding can slightly reduce the estimated source density in optical HST deep fields, though it is unlikely to be a large effect if one assumes a reasonable number density of galaxies in the Universe. The impact of crowding on galaxy detection will also depend on the effective wavelengths of the available filters, as the effective galaxy sizes also depend somewhat on the sampled rest-frame wavelengths. Simulations suggest that in some high-redshift galaxies, rest-frame optical light extends further into the galaxy outskirts than rest-frame ultraviolet light \citep[e.g.,][]{Ma2018}. JWST's near-infrared bands are able to trace the rest-frame optical light at the typical redshifts sampled (z$\simeq$1--3). Therefore, the extended rest-frame optical emission may start to overlap statistically to some extent with neighboring faint galaxies, making some individual galaxies harder to distinguish and complete samples harder to assess in the deep JWST images. For our current study, we will use the JWST NIRCam images in the JADES GOODS-S field of \cite{Eisenstein2026}.

In this study, we use NIRCam imaging from the JADES program to examine how increasing source density affects the recovered galaxy population and, by extension, the inferred completeness of faint source counts. Our goal is to test whether crowding in deep JWST data could reduce the IGL component of the EBL at infrared wavelengths. Section~\ref{sec:data} describes the JADES observations used in this study. Section~\ref{sec:methods} presents the construction of the replicated JADES and simulated MIRAGE images, the source-extraction procedure, and the diagnostics used to track how galaxy recovery changes as the fields become more crowded. Section~\ref{sec:results} presents the galaxy recovery fractions as a function of source density and magnitude. We then summarize our key findings in Section~\ref{sec:conclusions}.

\section{Data}
\label{sec:data}
This study uses publicly released JADES NIRCam drizzled v2.0 mosaics in the GOODS-S deep field \citep{Rieke2023}. For each filter, the input data consist of a science mosaic and its corresponding weight image. This particular field was chosen due to its available deep observations, wide near-infrared filter coverage and well-documented public data. The filters used in this analysis cover the near-infrared bands that are available in the JADES products: F090W, F115W, F150W, F200W,  F277W, F356W, and F444W. We list their exposure times in Table \ref{tab:exptimes}. Together, these filters sample a wide portion of the near-infrared range. This wavelength coverage is well suited for probing the Cosmic Infrared Background (CIB) and allows us to track how completeness evolves across the infrared regime. 

\begin{table}
\centering
\caption{JWST/NIRCam filters included in the JADES image analysis and their corresponding total exposure times.}
\label{tab:exptimes}
\begin{tabular}{lc}
\hline\hline
Filter & Exposure Time (s) \\
\hline
F090W & 62359 \\
F115W & 81170 \\
F150W & 73697 \\
F200W & 141725 \\
F277W & 19842 \\
F356W & 5669 \\
F444W & 24351 \\
\hline
\end{tabular}
\end{table}

\section{Methods}
\label{sec:methods}

Our analysis was designed to examine how increasing the projected galaxy density affects source recovery while keeping the observational characteristics of the data constant. Our analysis builds on the approach introduced by \citet{Kramer2022}, adjusting it to account for the pixel scale, sensitivity, and galaxy population of the JWST/NIRCam JADES mosaics. Using the observed JADES field, we increased the source population in three stages to produce the $2\times$, $3\times$, and $4\times$ realizations of the well-exposed inner square sub-image. For reasons we explain in detail below, we do not rotate each entire image successively by $90^{\circ}$ and add it to the previous image stack, unlike the procedure in \cite{Kramer2022}. Instead, each replicated individual source was taken directly from the observed image rather than being created separately, and then successively rotated by $90^{\circ}$ 
and added to the previous square image stack that contains the original image noise. This then allows us to quantitatively assess how large a fraction of faint extended objects is going to be lost due to statistical source overlap in the $2\times$, $3\times$, and $4\times$ rotated image realizations, which we take to represent hypothetical universes with $2\times$, $3\times$, and $4\times$ the actual source density. From their respective source catalogs and total galaxy counts we can then assess how large a fraction of known objects (from the original JADES HUDF images) would have been lost if the universe
and its IGL had really contained $2\times$, $3\times$, and $4\times$ the number of objects. 

For each of the JADES filters, we begin with matching cutouts from the science and weight images and use Source Extractor to identify the sources in each filter. Bright and compact objects are removed, and the remaining source population is placed into an object-only image. Rotated copies of the image are added to the masked JADES field, and the same source-extraction and catalog-selection procedures are applied to each realization. This method of ``pasting'' objects on top of the original, masked image was done to maintain the original image's background in the replications.

\subsection{Image Preparation}

For each filter, we use the drizzled JADES v2.0 science mosaic and its corresponding weight image \citep{Johnson2026,Robertson2026}, extracting the same $2700 \times 2700$ pixel cutout from both images to preserve their pixel-by-pixel alignment and simplify image rotation and addition. The cutout is centered at array coordinates $(16200,10550)$, corresponding to $\mathrm{RA}=03^{\mathrm{h}}32^{\mathrm{m}}19.80^{\mathrm{s}}$ and $\mathrm{Dec}=-27^{\circ}51^{\prime}25.74^{\prime\prime}$, or $(53.08248^{\circ},-27.85715^{\circ})$. At a pixel scale of $0.03^{\prime\prime}\,\mathrm{pixel}^{-1}$, the cutout measures $81^{\prime\prime}$ along each side, giving a total area of approximately $1.82\ \mathrm{arcmin}^{2}$.

After cropping, we update the World Coordinate System information to account for the cutout's position within the original mosaic and revise the FITS header keywords to reflect the dimensions of the cropped arrays. The output files retain the photometric calibration information from the original JADES mosaics, including the image units and pixel-area keywords.

\subsection{Source Detection and Mask Construction}

We identify and measure sources with Source Extractor \citep{BertinArnouts1996}, using the same detection settings throughout the analysis to generate masks, create object-only images, and compare the sources recovered from the original and replicated JADES fields.

We first identify and mask the sources that should be excluded from replication, then construct the replicated images from the retained source population, and finally measure the sources recovered from each realization.

\subsubsection{Bright-Object Identification and Masking}

During the first detection pass, Source Extractor processes the cropped science image together with its corresponding weight map, which we supply through the \texttt{MAP\_WEIGHT} configuration to account for differences in the statistical weight of individual pixels. To retain faint, compact sources while limiting detections caused by isolated noise fluctuations, we set both \texttt{DETECT\_THRESH} and \texttt{ANALYSIS\_THRESH} to 3.0 and set \texttt{DETECT\_MINAREA} to 4, which requires at least four connected pixels above the detection threshold.

Before sources are detected, the image is filtered with a $5\times5$ pixel Gaussian kernel with a full width at half maximum of 3 pixels. For deblending, we use 32 subthreshold levels and set \texttt{DEBLEND\_MINCONT}=0.03. These parameters control whether neighboring peaks within a shared detection region are treated as separate sources. The selected contrast separates nearby objects while limiting cases in which a single galaxy is divided into several catalog entries. Source Extractor estimates the background using a 64-pixel mesh and a three-mesh smoothing filter. Cleaning is enabled with \texttt{CLEAN\_PARAM}=1.0 to reduce false detections associated with brighter neighboring sources.

This first Source Extractor run creates both a source catalog and a segmentation image. Because objects in this field are more visibly extended than Source Extractor accounts for, we expand the segmented regions using \texttt{expand\_labels} from \texttt{scikit-image}. This process was manual; each image was masked with its radii increased until we considered the wings to be visibly removed. Filter-specific values are used because the NIRCam point-spread function and the measured apparent sizes of detected objects vary with wavelength and exposure time. The selected radii cover residual source light without removing an excessive portion of the image area.

The abundance of bright galaxies is well constrained by large ground-based surveys, so we are primarily concerned with the effect of confusion on faint sources. With that in mind, we only replicate faint sources as part of our procedure. Both magnitude and size are considered together to retain the faint galaxy population that is most relevant to the completeness analysis. Objects selected for removal are identified by their positions in the \texttt{MAG\_AUTO} versus \texttt{FLUX\_RADIUS} distribution. \texttt{MAG\_AUTO} is a Kron-like estimate of the total source magnitude. Sources with $\texttt{MAG\_AUTO}<24$ mag are removed so that bright objects and their extended wings are not present in the replicated fields. A second selection removes sources with radii below the filter-specific radius limit, $r_{\mathrm{cut}}$, which is determined from the compact stellar locus; any detections smaller than point-like stars are not real objects. Objects meeting either condition are selected for removal. The population retained for replication therefore consists primarily of faint, resolved sources used in the galaxy-count completeness test. The filter-specific mask-expansion radii and half-light-radius cuts are listed in Table~\ref{tab:mask_radii}.

\begin{deluxetable}{lcccc}
\tabletypesize{\scriptsize}
\tablecaption{Segmentation-mask expansion radii and half-light-radius
cuts adopted for each JADES filter.\label{tab:mask_radii}}
\tablehead{
\colhead{Filter} &
\colhead{$r_{\mathrm{exp}}$ (pix)} &
\colhead{$r_{\mathrm{exp}}$ (arcsec)} &
\colhead{$r_{\mathrm{cut}}$ (pix)} &
\colhead{$r_{\mathrm{cut}}$ (arcsec)}
}
\startdata
F090W & 24 & 0.72 & 1.25 & 0.038 \\
F115W & 24 & 0.72 & 1.50 & 0.045 \\
F150W & 24 & 0.72 & 1.35 & 0.041 \\
F200W & 30 & 0.90 & 1.25 & 0.038 \\
F277W & 24 & 0.72 & 2.00 & 0.060 \\
F356W & 24 & 0.72 & 2.50 & 0.075 \\
F444W & 24 & 0.72 & 2.75 & 0.083 \\
\enddata
\tablecomments{Angular values were calculated using the JADES pixel scale of $0.03^{\prime\prime}$ pixel$^{-1}$.}
\end{deluxetable}

Inspection of the masked images showed that a 24-pixel expansion radius was sufficient for all filters except F200W. In F200W, faint light from the outer wings of bright galaxies remained visible beyond 24 pixels, so the expansion radius was increased to 30 pixels. At the JADES pixel scale of $0\farcs03\,\mathrm{pixel}^{-1}$, these radii correspond to $0\farcs72$ and $0\farcs90$, respectively. These values cover the residual source light while limiting the amount of additional background area included in the masks.

The selected pixels are filled with background-like noise rather than being set to zero. Setting these pixels to zero would create empty regions with abrupt boundaries that do not resemble the surrounding image. Instead, the pixels are replaced with random values drawn from a Gaussian distribution. The mean and standard deviation of this distribution are taken from the background and background RMS measured by Source Extractor. A fixed random seed of 42 makes the replacement reproducible each time the code is run. The final masked image removes bright and compact sources while keeping the statistical properties of the background approximately consistent with those of the unmasked regions. Once the selected sources have been replaced with background-like noise, the masked image is used to construct the replicated fields.

\subsubsection{Construction of Replicated Fields}

A second Source Extractor pass is run on the masked science image with the corresponding weight map. For this step, \texttt{CHECKIMAGE\_TYPE} is set to \texttt{OBJECTS}. The resulting check image contains the measured flux from detected sources but does not include the surrounding sky background. This differs from the segmentation map, where the pixels in each source are represented by an integer label instead of their measured flux.

The object-only image is used to increase the source density without also duplicating the background. If the full science image were rotated and added to itself, the sky background and correlated noise would be repeated along with the sources. This would change both the source density and the effective depth of the image. We instead retain the original masked science image's background and noise realization to the best of our ability, then add rotated copies of only the detected source population.

The object-only image was rotated by $90^{\circ}$, $180^{\circ}$, and $270^{\circ}$. Due to the square shape of the cutout, every copy remains on the original pixel grid. We produced the quarter-turn rotations by transposing the array and then flipping it along one axis, with the choice of the axis determining the direction of rotation. For the half turn, we reversed both of the axes. Each operation moved each pixel directly without interpolation, preserving the sampled source profiles. 

Let $I_0$ denote the masked JADES science image, and let $O_{\theta}$ denote the object-only image after a rotation through an angle $\theta$. We construct the replicated images as

\begin{align}
I_2 &= I_0 + O_{90}, \\
I_3 &= I_0 + O_{90} + O_{180}, \\
I_4 &= I_0 + O_{90} + O_{180} + O_{270}.
\end{align}

We refer to these images as the $2\times$, $3\times$, and $4\times$ realizations. These labels describe the number of copies of the retained source population that were placed in each field. They do not imply that Source Extractor will recover exactly two, three, or four times the original number of detections. As the field becomes more crowded, some sources overlap or blend together and may be recorded as a single object. Other sources may no longer satisfy the detection requirements. The difference between the number of sources added and the number recovered is used to measure completeness.

The weight map is rotated by the same angles used for the object-only image. The JADES weight maps are expressed as inverse variance. We first convert each component to variance, add the variances, and then convert the combined result back to inverse variance. For a realization containing $n$ image components, the combined weight is

\begin{equation}
W_n(x,y)=
\left[
\sum_{k=1}^{n}\frac{1}{W_k(x,y)}
\right]^{-1},
\end{equation}

\noindent where $W_k(x,y)$ is the original or rotated weight map associated with component $k$. Any pixel with an undefined, infinite, or negative combined weight is assigned a value of zero. The replicated science images and their corresponding weight maps are then saved as matched FITS files using the headers inherited from the cropped JADES products.

\subsubsection{Catalog Construction and Completeness Measurement}

The purpose of this comparison is to determine how increasing source crowding affects source recovery in the JADES fields. Any differences indicate that some sources are no longer being recovered as separate detections. Source Extractor is run separately on the final masked JADES image and on the $2\times$, $3\times$, and $4\times$ images. The detection threshold, minimum detection area, deblending contrast, and treatment of the weight map remain the same for every run. Holding these settings fixed allows us to compare the catalogs directly. Any changes in the recovered source counts can then be associated with the increasing source density rather than a change in the detection procedure.

We select sources from each catalog using the same magnitude, radius, and flag requirements. The retained sample satisfies \[ \texttt{MAG\_AUTO}>24
\quad\text{and}\quad \texttt{FLUX\_RADIUS}>r_{\mathrm{cut}}. \] We also remove catalog entries with \[ \texttt{FLAGS}\geq4. \]

Detections with $\texttt{FLAGS}\geq4$ are automatically excluded because their photometry or morphology may be unreliable. Using these criteria throughout this analysis keeps the selected source population consistent across the replicated catalogs.

Differential source counts are calculated in 0.5 magnitude bins for each filter. The counts are normalized by the angular size of the image cutout, and the Poisson uncertainty in each bin is calculated from $\sqrt{N}$. In an ideal $n\times$ image, the number of sources would equal $n$ times the number measured in the original masked JADES image. We therefore define the magnitude-dependent recovery fraction as
\begin{equation}
C_n(m)=
\frac{N_{\mathrm{recovered},n}(m)}
     {nN_{\mathrm{JADES}}(m)},
\end{equation}

\noindent where $N_{\mathrm{recovered},n}(m)$ is the number of sources detected in the replicated image within magnitude bin $m$, and $N_{\mathrm{JADES}}(m)$ is the corresponding number from the original masked field. A recovery fraction of $C_n(m)=1$ means that the measured count agrees with the number expected from simple replication. A value below one means that some of the added sources were not recovered as separate detections. Uncertainties in the recovery fractions are propagated from the Poisson uncertainties in the measured source counts. Spline curves are shown in the figures to make the overall trends easier to see, but the completeness measurements are based on the binned counts.

\begin{figure*}[!t]
    \centering
    \includegraphics[width=0.60\textwidth]{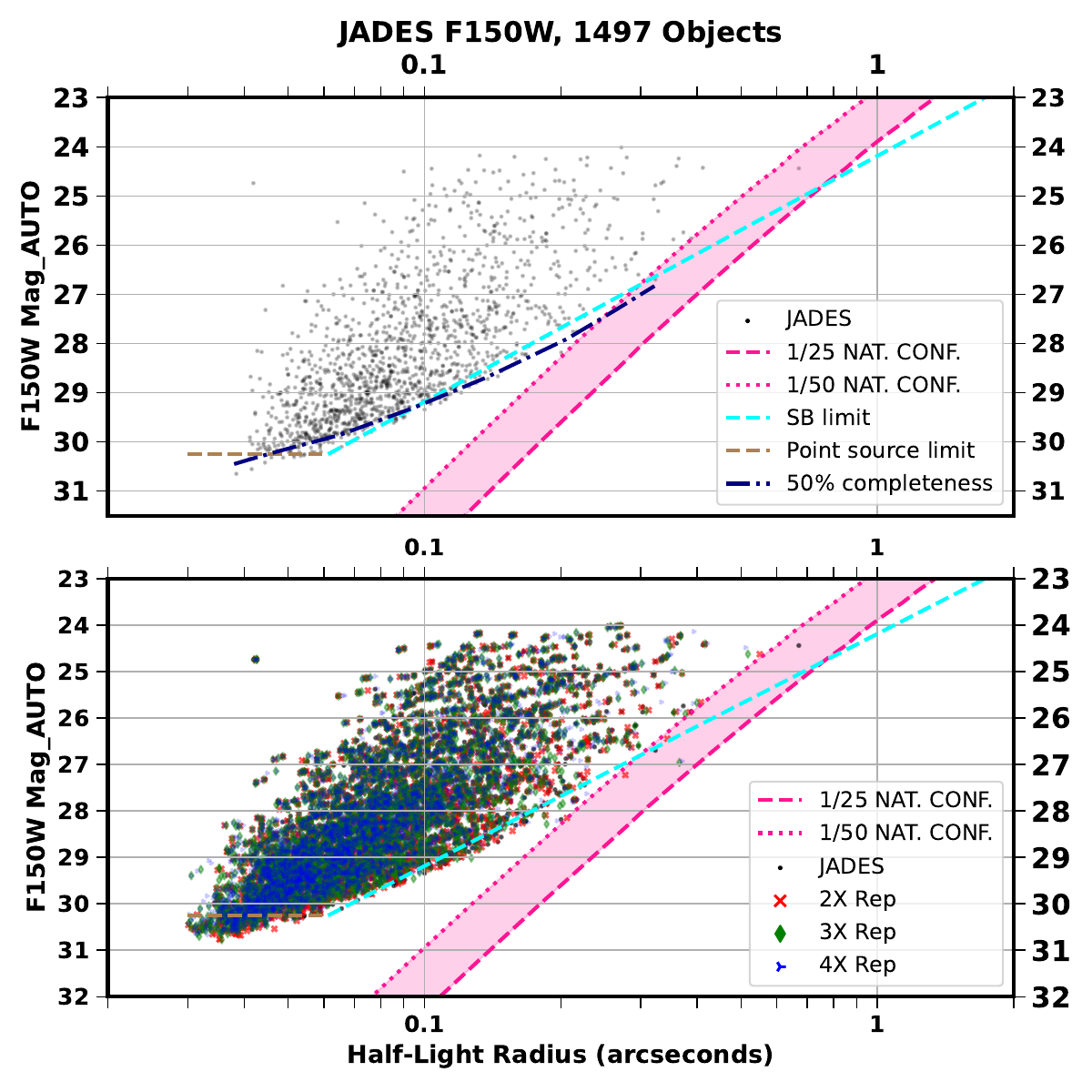}
    \caption{Magnitude versus half-light radius for the JADES F150W catalog. The upper panel shows the original JADES source distribution relative to the point-source limit (brown line), the surface-brightness detection limit (cyan line), and the natural confusion regime defined by 1/25 to 1/50 objects per beam (pink dashed and dotted lines, respectively). The navy dash-dotted curve shows the 50\% completeness limit measured from injected sources (Section~\ref{sec:methods}). The lower panel includes the $2\times$, $3\times$, and $4\times$ replicated realizations, demonstrating how increasing source density drives more objects into the region where confusion and size-dependent incompleteness become important.}
    \label{fig:f150w_magnitude_size}
\end{figure*}

We also measured the 50\% completeness limit of the F150W catalog with the injection-recovery pipeline of \citet{Oskouie2026}, injecting 200 exponential-disk galaxies per magnitude and size into the masked image and detecting them with the same Source Extractor settings used for the JADES catalogs. To place the limit on the axes of Figure~\ref{fig:f150w_magnitude_size}, we convert each 50\% completeness magnitude and effective radius to the median \texttt{MAG\_AUTO} and \texttt{FLUX\_RADIUS} of recovered sources injected within 0.5 mag of that limit.

\section{Results}\label{sec:results}

The replicated JADES images show that source recovery falls increasingly below the counts expected from simple replication as source density rises. The number of recovered detections increases as more objects are added, but it does not increase at the rate expected if every added source were detected separately. The difference becomes larger in the more crowded images and is most noticeable at faint magnitudes.

\subsection{Recovery in the Replicated JADES Images}

Figure~\ref{fig:f150w_magnitude_size} shows the magnitude and half-light-radius distributions for the F150W catalog. The upper panel contains the 1497 sources measured in the original JADES image. The point-source limit, surface-brightness boundary, and natural-confusion range are included as references. Faint detections are concentrated at smaller half-light radii, while brighter objects cover a wider range of sizes.

The 50\% completeness curve agrees with the reference lines measured independently from the catalog. At the smallest radii it reaches 30.45 mag, close to the point-source limit, although it is steeper there because Source Extractor measures injected sizes from $0\farcs01$ to $0\farcs056$ at nearly the same radius near the detection limit. At intermediate radii the curve follows the faint edge of the JADES source distribution, and at its largest radii it falls on the surface-brightness limit. The curve ends at a measured radius of about $0\farcs32$, since near the detection limit Source Extractor recovers only the cores of larger galaxies. At that radius it reaches the edge of the natural confusion regime, so faint, extended galaxies near the detection limit are also the ones most likely to be affected by crowding.

The lower panel includes the original catalog and the catalogs recovered from the $2\times$, $3\times$, and $4\times$ images. Since the added sources were taken from the original JADES field, the catalogs remain within a similar region of the magnitude-size plane. Increasing the replication factor mainly adds detections to this existing distribution. The faintest sources lie closest to the surface-brightness and natural-confusion boundaries, making this part of the catalog more likely to be affected by crowding.

Figure~\ref{fig:f150w_completeness_summary} provides the count-based measurement of this effect. The source density measured in the original F150W catalog is $3.21 \times 10^{6}$ objects deg$^{-2}$. The recovered densities increase to $5.08 \times 10^{6}$, $6.87 \times 10^{6}$, and $8.51 \times 10^{6}$ objects deg$^{-2}$ in the $2\times$, $3\times$, and $4\times$ images, respectively. Although the measured density rises with each replication, it remains below the value expected from directly multiplying the original count.

Compared with the counts predicted by a simple replication, the $2\times$, $3\times$, and $4\times$ catalogs recover $79.0\%$, $71.3\%$, and $66.2\%$ of their expected populations. The fraction missed rises from $21\%$ in the $2\times$ field to $29\%$ in the $3\times$ field and $34\%$ in the $4\times$ field, respectively. The growing difference indicates that a larger fraction of the added sources can no longer be identified as distinct sources at higher densities.

The right panel of Figure~\ref{fig:f150w_completeness_summary} shows the recovery ratios by their magnitude. Brighter bins remain consistent with a ratio of one within their uncertainties, but they decrease towards the faint end. The $4\times$ realization has the lowest ratios generally. This pattern points to a gradual decline in recovery toward fainter magnitudes rather than a single, sharply defined confusion threshold.

\subsection{Comparison Across JADES Filters}
Figure~\ref{fig:mag_rhalf_multi} compares the magnitude-size distributions in F115W, F200W, and F444W. The original catalogs contain 1781, 1762, and 1290 sources, respectively. In each filter, the replicated catalogs remain close to the distribution measured in the original JADES field. The number of detections within that distribution increases from the $2\times$ to the $4\times$ image, while the faint sources remain nearest to the surface brightness and natural confusion boundaries.

The location of the source distribution changes with wavelength. The F444W detections generally extend to larger half-light radii than the sources in F115W and F200W. This change is consistent with the broader NIRCam point-spread function at longer wavelengths. It also changes the position of the catalog relative to the size-dependent reference curves. The influence of crowding is therefore not identical in every filter, even though the replicated catalogs follow a similar overall pattern.

Figure~\ref{fig:jades_multifilter_completeness} compares the differential source counts and recovery ratios in F150W, F200W, and F444W.

\section{Conclusion}\label{sec:conclusions}

We used deep JWST NIRCam observations from JADES to examine how source crowding affects the detection of faint galaxies, creating progressively denser versions of the field by adding rotated copies of detected sources to the masked image. Relative to the counts predicted by scaling the original F150W catalog, the fraction of sources recovered fell from 79.0\% in the $2\times$ realization to 71.3\% in the $3\times$ realization, and to 66.2\% in the $4\times$ realization, respectively. The deficit was most pronounced at faint magnitudes, where galaxies were less likely to be recovered as distinct sources. 

This decline in recovery matters because deep galaxy counts are used to estimate the amount of light that resolved galaxies contribute to the CIB. When neighboring galaxies blend into a single detection, their combined light can be counted in a brighter magnitude bin even if much of their flux is still measured. Assessing the impact of crowding on integrated galaxy light therefore requires measuring both how much flux is recovered from blended sources and how many galaxies are detected.

From the above numbers, one might conclude that between 21--34\% of the IGL would have been missed if the universe had been 2--4$\times$ more dense with galaxies than it currently is. However, as discussed in \cite{Carleton2022}, \cite{Windhorst2022}, and \cite{Windhorst2023}, most of the IGL comes from galaxies with AB$\simeq$18--24 mag with the IGL contribution peaking at AB$\simeq$22 mag, and less than 2.5--3\% of the extrapolated IGL coming from galaxies with AB$\gtrsim$27--28 mag. Therefore, even if 21--34\% of the galaxies with AB$\gtrsim$29 mag had been missed in the deepest JWST NIRCam images, the total IGL (including the missing part coming from AB$\gtrsim$29 mag) would have been underestimated by far less, likely by less than 1\%. Hence, we consider it unlikely that any Diffuse Light is coming from faint galaxy populations not detected in the deepest JWST images due to source overlap, and its causes need to be found elsewhere, e.g., in a missing dim, diffuse Zodiacal component in the inner solar system that was not present in previous models \citep[e.g.,][]{OBrien2026, Windhorst2023}. 

A next step is to repeat these tests in independent deep JWST fields and measure how crowding affects both source counts and recovered flux across the NIRCam filters. Comparing these measurements could help establish how recovery varies with wavelength and source density and quantify how much crowding affects estimates of integrated galaxy light in the near-infrared.

\onecolumngrid\vspace{6pt}
\noindent
\begin{minipage}{\textwidth}
    \centering
    \makeatletter
    \def\@captype{figure}
    \makeatother
    \includegraphics[width=\linewidth]{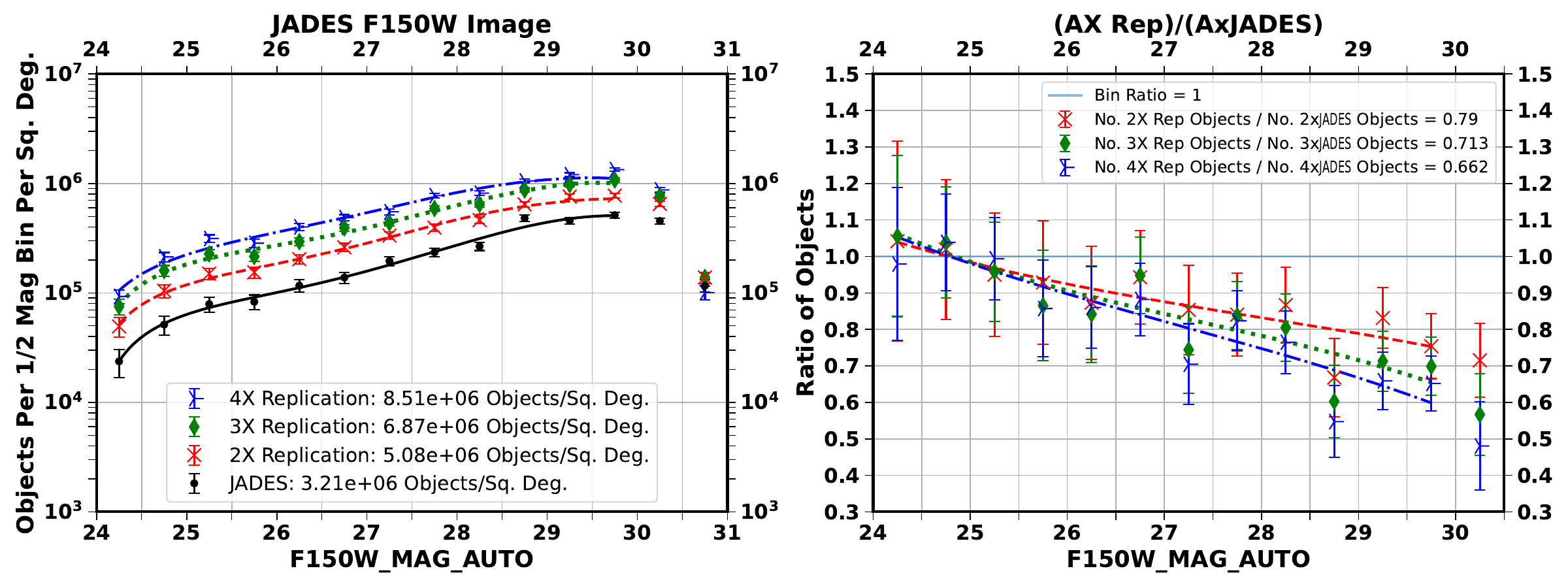}
    \caption{Source recovery in the JADES F150W field. The left panel compares differential source counts per square degree in the 0.5 mag bins for the original catalog and the $2\times$, $3\times$, and $4\times$ realizations. The right panel shows the ratio of the measured count in each bin to the count expected from scaling the original catalog by the replication factor. The horizontal line at unity marks where the recovered and expected counts agree. The shortfall grows towards fainter magnitudes and is generally largest in the $4\times$ realization.}
    \label{fig:f150w_completeness_summary}
\end{minipage}
\par

\clearpage
\noindent
\begin{minipage}{\textwidth}
    \centering
    \makeatletter
    \def\@captype{figure}
    \makeatother
    \includegraphics[width=0.3333\linewidth]{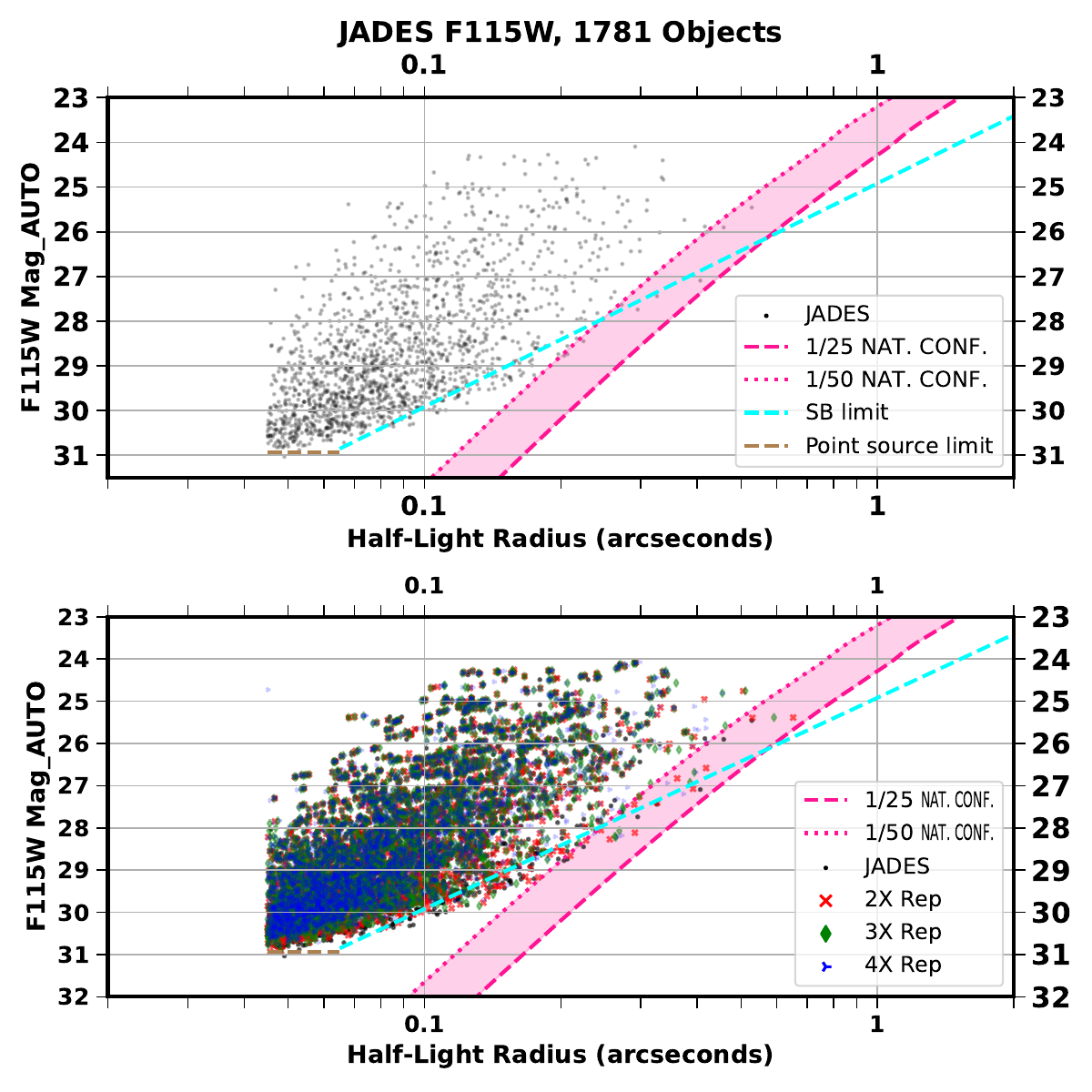}%
    \includegraphics[width=0.3333\linewidth]{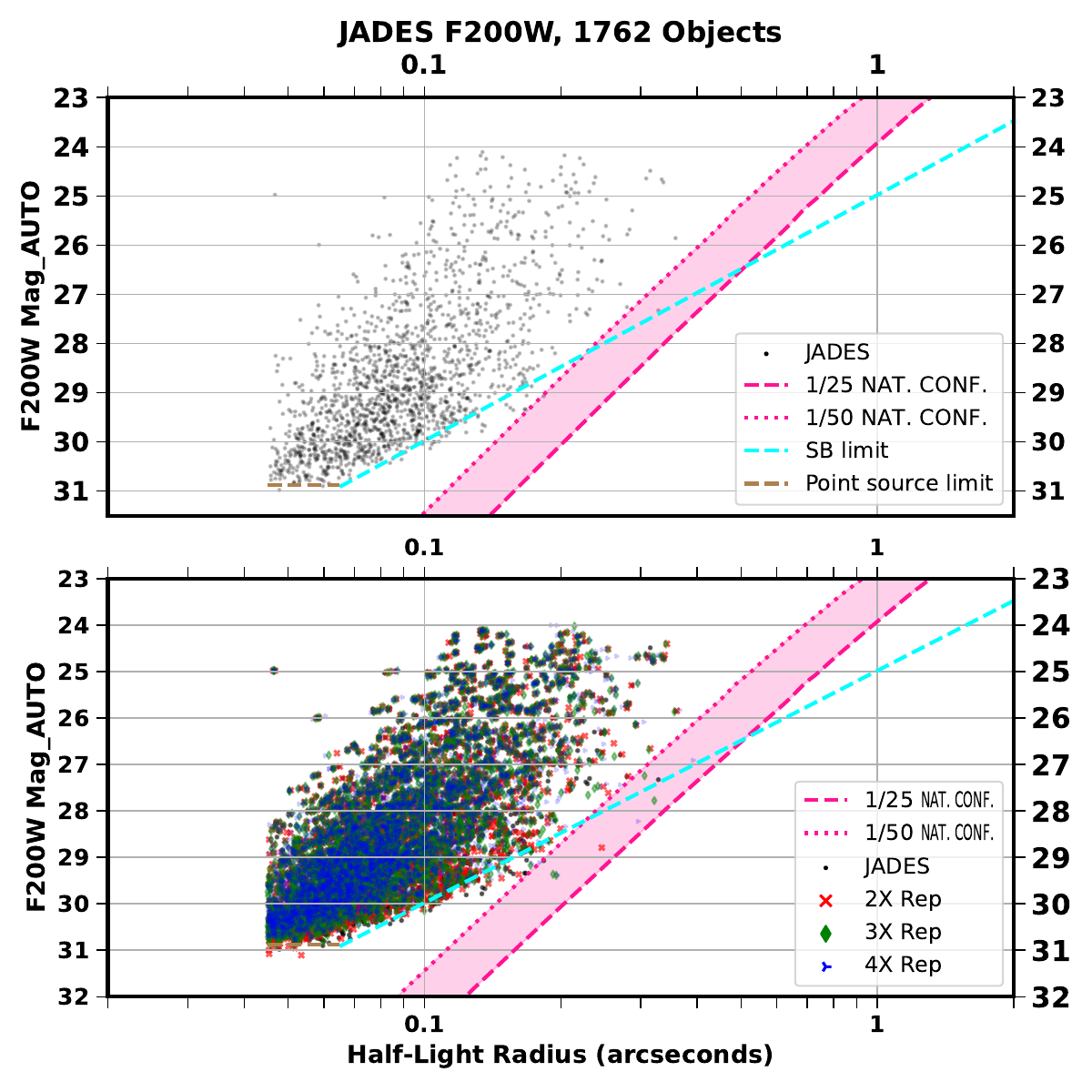}%
    \includegraphics[width=0.3333\linewidth]{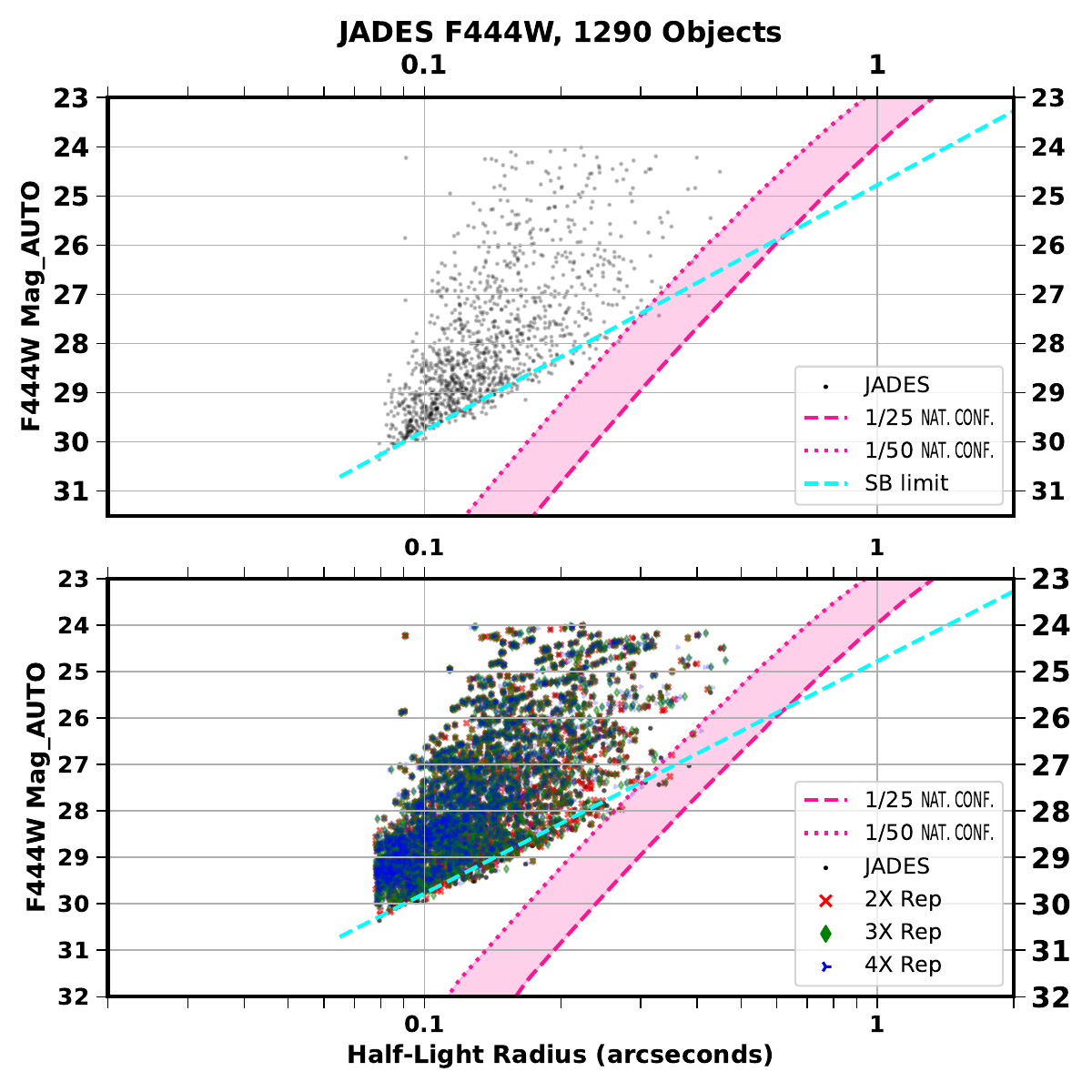}%
    \caption{
    Magnitude versus half-light radius across representative JADES filters F115W, F200W, and F444W. The upper row shows the original JADES catalogs, and the lower row shows the corresponding $2\times$, $3\times$, and $4\times$ replicated realizations. The pink band marks the natural confusion range between 1/25 and 1/50 objects per beam, while the cyan dashed curve indicates the surface-brightness detection boundary. The progression from shorter to longer wavelength illustrates how the balance between angular resolution, source size, and crowding changes across the dataset and affects the completeness of faint galaxy recovery.}
    \label{fig:mag_rhalf_multi}
    \par\medskip
    \includegraphics[width=\textwidth]{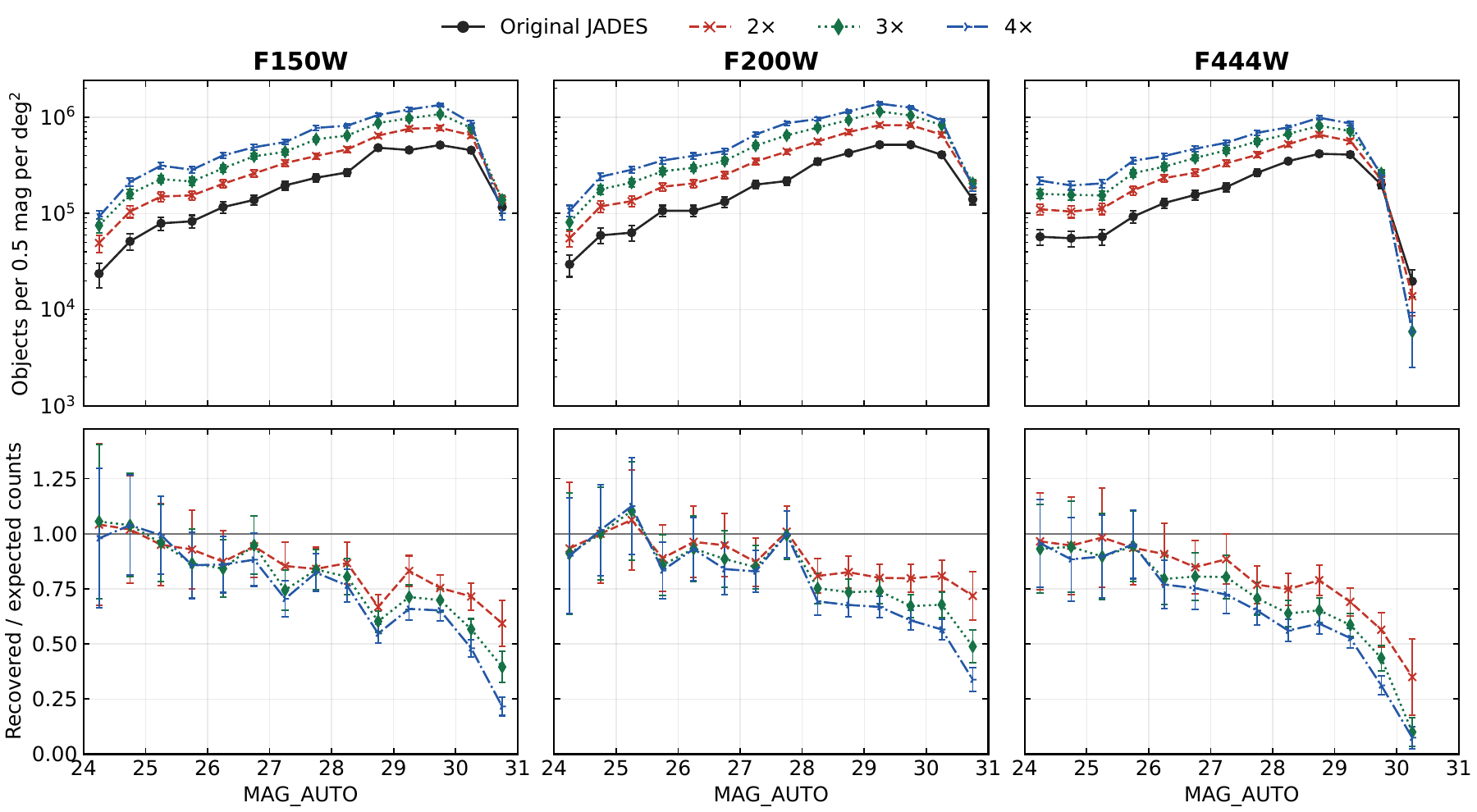}
    \caption{Source counts and recovery ratios for JADES F150W, F200W, and F444W from left to right. The upper panels show the number of sources per square degree in 0.5 mag bins for the original JADES field and the $2\times$, $3\times$, and $4\times$ realizations. The lower panels show the recovered counts divided by the counts expected from multiplying the original counts in each bin by the replication factor. The horizontal line at unity marks the expected ratio for full recovery, and the error bars show Poisson uncertainties in the counts and their propagation into the ratios. Recovery falls furthest below unity towards fainter magnitudes, particularly in the $4\times$ realizations.}
    \label{fig:jades_multifilter_completeness}
\end{minipage}
\par

\clearpage
\bibliography{ref}{}
\bibliographystyle{aasjournalv7}

\end{document}